# Magnetic Superstructure in the Two-Dimensional Quantum Antiferromagnet $SrCu_2(BO_3)_2$

K. Kodama,[1] M. Takigawa,[1]* M. Horvatić,[2] C. Berthier,[2,3] H. Kageyama,[1] Y. Ueda,[1]
S. Miyahara,[1,4] F. Becca,[4] F. Mila[4]

[1]*Institute for Solid State Physics, University of Tokyo, Kashiwa, Chiba 277-8581, Japan*
[2]*Grenoble High Magnetic Field Laboratory, CNRS and MPI-FKF, 38042 Grenoble, France*
[3]*Laboratoire de Spectrométrie Physique, Université Joseph Fourier Grenoble I, 38402 St.-Martin d'Hères, France*
[4]*Institut de Physique Théorique, Université de Lausannne, CH-1015 Lausanne, Switzerland*
*To whom correspondence should be addressed. E-mail:masashi@issp.u-tokyo.ac.jp

Abstract

We report the observation of magnetic superstructure in a magnetization plateau state of $SrCu_2(BO_3)_2$, a frustrated quasi-two-dimensional quantum spin system. The Cu and B nuclear magnetic resonance (NMR) spectra at 35 millikelvin indicate an apparently discontinuous phase transition from uniform magnetization to a modulated superstructure near 27 tesla, above which a magnetization plateau at 1/8 of the full saturation has been observed. Comparison of the Cu NMR spectrum and the theoretical analysis of a Heisenberg spin model demonstrates the crystallization of itinerant triplets in the plateau phase within a large rhomboid unit cell (16 spins per layer) showing oscillations of the spin polarization. Thus we are now in possession of an interesting model system to study a localization transition of strongly interacting quantum particles.

The competition between itinerancy favored by kinetic energy and localization favored by repulsive interactions is a fundamental aspect of many-body quantum systems, of which the Mott (metal-insulator) transition is an example (*1*). Similar phenomena may occur for the spin degrees of freedom of certain magnetic insulators, known as spin-liquids, in which the ground state is a singlet separated from triplet excitations by a finite energy gap $\Delta$. Transition metal oxides exhibiting this kind of physics have been actively studied in an effort to understand the more complex physics of high-temperature superconductors [e.g., (*2*)], a good example being spin ladders (*3*). In such systems, a magnetic field $H_c = \Delta/g\mu_B$ (where g is the effective g-value of the electron spin and $\mu_B$ is the Bohr magneton) will close the gap and the resulting magnetization is a gas of mobile triplets whose density can be tuned by the field value. However, when this density becomes commensurate with the underlying crystal lattice, the triplets may crystallize into a superlattice. The magnetization density will then stay constant for a certain range of magnetic field resulting in a magnetization plateau (*4, 5*), which has indeed been observed in some quantum spin systems (*6, 7*). It was argued theoretically that a magnetization plateau at a fractional value of $1\mu_B$ per unit cell occurs only when translational symmetry of the crystal lattice is broken by a magnetic superlattice (*8*). We report direct observation of such a magnetic superlattice by NMR experiments in $SrCu_2(BO_3)_2$, a quasi-two-dimensional quantum spin system.

The planar network of orthogonal dimers of spin 1/2 $Cu^{2+}$ ions in $SrCu_2(BO_3)_2$ (Fig. 1A) has stimulated extensive experimental (*10-16*) and theoretical (*17-23*) work. This compound is a realization of the Shastry-Sutherland model (Fig. 1B) conceived theoretically two decades ago (*24*):

$$H = J \sum_{(n.n.)} \mathbf{S}_i \cdot \mathbf{S}_j + J' \sum_{(n.n.n.)} \mathbf{S}_i \cdot \mathbf{S}_j \quad (1),$$

where $J$ and $J'$ are respectively the nearest neighbor (intra-dimer) and the next nearest neighbor (inter-dimer) antiferromagnetic exchange constants between two $S=1/2$ spins $\mathbf{S}_i$ and $\mathbf{S}_j$. When $J$ is much larger than $J'$, the model reduces to a collection of dimers. In the opposite limit, $J/J'<<1$, it is equivalent to the antiferromagnetic Heisenberg model on a square lattice with a Néel ordered ground state. The Shastry-Sutherland model has the property that the direct-product of the dimer singlet states is the exact ground state when $J'/J$ is smaller than a certain critical value near 0.7 (*17, 18, 22, 24*). An excited state can be made by promoting a singlet dimer into a triplet state, which can then hop from one dimer to another when $J' \neq 0$. The frustration between $J$ and $J'$ in the Shastry-Sutherland model, however, inhibits such hopping up to fifth order of the perturbation series in $J'/J$, leading to extremely flat triplet bands, i.e. a very small kinetic energy (*18*).

Various experiments have established that the dimer singlet ground state (*10, 12*) is indeed realized in $SrCu_2(BO_3)_2$, with an energy gap to magnetic excitations of 3.0 meV (*13, 14*). Nearly flat triplet bands were also observed by neutron inelastic scattering experiments (*14, 15*). The most striking property of $SrCu_2(BO_3)_2$ is the plateaus in the magnetization curve at 1/8, 1/4, and 1/3 of the fully saturated moment (*10, 11*). One can expect that the small kinetic energy of triplets in the Shastry-Sutherland model will allow them to crystallize into a superlattice at these commensurate densities if there are repulsive interactions between triplets. This is analogous to the Wigner crystallization or charge ordering of electron systems. Although theories have indeed supported such a picture (*19-21, 23*) and predicted possible structures of the superlattices (*19-21*), no experimental evidence has yet been reported. Inequivalent Cu sites in a commensurate magnetic superstructure should have different hyperfine fields that manifest as distinct NMR lines. Therefore, we performed NMR measurements on a single crystal of $SrCu_2(BO_3)_2$ grown by the traveling-solvent-floating-zone method (*25*) in a magnetic field up to 28 T applied along the *c*-axis at 35 mK. This covers the first "1/8" magnetization plateau which has been observed (*11*) in the field range 27 to 28.5 T.

We show in Fig. 2 the Cu NMR spectra in a field of 26 T (inset) and 27.6 T (main panel). A single Cu site gives six NMR lines at the following frequencies (*26*),

$$f(\mathbf{a},m) = {}^{a}g\{(1+K)H_0 + H_n\} + {}^{a}n_c(m-1/2) \quad (2).$$

The first term is the Zeeman frequency due to the external field $H_0$ corrected for the orbital shift $K$= 1.69 % (*12*) and the hyperfine field $H_n$, ${}^{\alpha}\gamma$ ($\alpha$=63 or 65) is the known gyromagnetic ratio of either $^{63}$Cu or $^{65}$Cu nuclei. In general, the spin polarizations both on the same site and on the neighboring sites contribute to $H_n$. Here we assume only the dominant on-site coupling, $H_n = A_c g_c \langle S_z \rangle$, where $\langle S_z \rangle$ is the time-averaged local magnetization. The coupling constant and the *g*-value were previously determined as $A_c = -23.8$ T/$\mu_B$ and $g_c = 2.28$ (*12, 13*). The second term represents the quadrupolar shift for spin 3/2 nuclei, where $m$=3/2, 1/2, or −1/2 distinguishes three transitions ($I_z = m \rightarrow m-1$) and $^{63}\nu_c$=22.1 MHz, $^{65}\nu_c$=20.5 MHz as previously determined (*12*). The Cu NMR spectrum at 26 T can be well fitted by Eq. 2 with a Gaussian distribution of $H_n$ peaked at –1.79 T with a full width at half maximum (FWHM) of 0.59 T, demonstrating that the system at 26 T is in the uniform phase ($\langle S_z \rangle$ = 0.033) where triplets are itinerant.

At a field of 27.6 T, a drastic change of the spectrum was observed with the appearance of many sharp peaks distributed over a wide frequency range, indicative of a commensurate magnetic order with a large unit cell. This is the first clear evidence for a magnetic superlattice in the 1/8 plateau phase with broken translational symmetry. Although the spectrum appears complicated, some important features are immediately recognized. The sharp six peaks at low frequencies, shown by the red zone in Fig. 2, have the intensity approximately 1/8 of the entire spectrum and their positions are given by Eq. 2 with $\langle S_z \rangle$ = 0.30. Likewise, the broad lines in the yellow zone in Fig. 2 can be ascribed to a single site with 1/8 population and $\langle S_z \rangle$ = 0.20. If the triplets were confined on a single dimer, 1/8 of the Cu sites would have $\langle S_z \rangle$ = 0.50 and $\langle S_z \rangle$ = 0 for the rest. Therefore, the magnetization must be spread over several sites within the magnetic unit cell. We tried to fit the entire spectrum by progressively adding a new set of lines given by Eq. 2 with a Lorentzian distribution of $H_n$ until all the peaks and the overall shape were reproduced. We found that at least 11 distinct Cu sites had to be included. The population of each site can be chosen to be an integral multiple of 1/16, which is compatible with a magnetic unit cell containing 16 spins. A successful fit is shown by the red line in Fig. 2 and the corresponding parameters are listed in the Table S1. The histogram of the peak values of $H_n$ is shown in the middle panel of Fig. 3B. Several sites have positive $H_n$ (blue zone in Fig. 2), implying (through a negative hyperfine coupling) that spins are polarized opposite to the magnetic field direction. Thus the magnetization oscillates within the unit cell. We confirmed that the average of $H_n$ over all the sites gives the spatially-averaged magnetization $\langle S_z \rangle$ = 0.063, exactly 1/8 of the saturation.

Let us examine whether our results are compatible with the Shastry-Sutherland model. In the simplest theoretical approach, interactions between two triplet dimers are calculated by a perturbation in $J'/J$ (*19-21*). Two different configurations (Fig. 1C and D) have been proposed for the 1/8 plateau (*21*). Here, each triplet is completely localized on one dimer, the result of an oversimplification of this approach. The real magnetization of the Shastry-Sutherland model spreads over all the sites as explained below. The number of inequivalent Cu sites, however, does not depend on such details of the magnetization profile and can be determined only from symmetry considerations. For the square cell, there are six distinct Cu sites, as shown in Fig. 1C. The number of Cu sites may increase depending on how the layers stack along the *c*-axis, however, only up to eight, which is clearly not sufficient to reproduce all the observed peaks. On the other hand, the rhomboid cell has eight Cu sites for a single layer (Fig. 1D) and the stacking of layers may increase the number up to 16, although all the sites may not be resolved experimentally. Therefore, we may expect that only the rhomboid cell is compatible with the experiments.

To go beyond these qualitative considerations and check whether the Shastry-Sutherland model can indeed reproduce the peculiar spin texture revealed by NMR, with positive and negative magnetizations, we must solve the original spin Hamiltonian of Eq. 1 for a finite size lattice. There is a subtlety though. If the ground state breaks the translational symmetry, it is expected to be eight-fold degenerate for an infinite system and the ground state magnetization is not well defined. In the actual compound, a unique ground state would be selected by pinning from impurities or from a lattice distortion. In our finite size calculation, although periodic boundary condition partially lifts the eight-fold degeneracy, some mechanism still must be added to the model to enable selection of one of the ground states (*27*). Because the pronounced softening of the sound velocity observed at the edges of the magnetization plateaus (*16*) suggests that a lattice distortion indeed occurs in SrCu$_2$(BO$_3$)$_2$, we decided to include an adiabatic spin-phonon coupling (*28*),

$$H = \sum_{(n.n.)} J\left(1 - a\frac{dd_{ij}}{d^0_{ij}}\right) \mathbf{S}_i \cdot \mathbf{S}_j + \sum_{(n.n.n.)} J'\left(1 - a'\frac{dd_{ij}}{d^0_{ij}}\right) \mathbf{S}_i \cdot \mathbf{S}_j$$
$$+ \frac{K}{2} \sum_{(n.n.)} \left(\frac{\|\mathbf{dr}_i - \mathbf{dr}_j\|}{d^0_{ij}}\right)^2 + \frac{K'}{2} \sum_{(n.n.n.)} \left(\frac{\|\mathbf{dr}_i - \mathbf{dr}_j\|}{d^0_{ij}}\right)^2 \quad (3)$$

parameterized by the spin-phonon coupling constants *a* and *a'* and the elastic constants *K* and *K'*. The exchange constants now depend on the deviation of the interatomic distance $dd_{ij}$ from the value $d^0_{ij}$ in the uniform phase. The atomic displacements $\mathbf{dr}_i$ are free to adjust to give the best compromise between the increase of elastic energy and the gain in magnetic energy. The Hamiltonian of Eq. 3 was solved by exact diagonalization of a 16-spin (eight-dimer) cluster using the Lanczos algorithm imposing periodic boundary conditions compatible with the rhomboid cell and a total magnetization of 1/8, and by minimizing the energy with respect to $\mathbf{dr}_i$. Similar calculations have been performed in the context of spin Peierls systems, such as CuGeO$_3$ (*29*).

The results of Fig. 3A were obtained assuming typical values *a*=*a'*=7 and *K*=*K'*=20000 K for oxides (*27*) and using well accepted values for the exchange constants, *J*=85 K and *J'*=54 K (*30*). The magnetization profile extends over the entire unit cell with one strongly polarized dimer surrounded by decaying oscillation of magnetization. Similar structure has also been observed around impurities in quantum spin chains (*31, 32*) and two-dimensional superconducting cuprates (*33*). It is analogous to the Friedel oscillation near impurities in metals. The calculated profile captures the essential feature of the experimental results (Fig. 3B), providing further support for the rhomboid cell. Note that the coupling to phonons is required merely to select one of the degenerate ground states and is not necessary to obtain a plateau. The magnetization pattern depends only weakly on the value of the spin-phonon coupling (Fig. S1), therefore, the qualitative feature mentioned above is expected to be an intrinsic aspect of the magnetic system. Note also that the displacements we obtain (Table S2) are very small – less than 1 %, which would not be possible to detect experimentally in a magnetic field of 27 T with facilities currently available.

We now discuss the nature of the phase transition into the 1/8 plateau state. The field dependence of the magnetization obtained from the Cu NMR spectra (Fig. 4A) shows a nearly discontinuous jump upon entering the 1/8 phase. The vicinity of the phase boundary was studied by the $^{11}$B NMR spectra obtained at 35 mK (Fig. 4B). The $^{11}$B spectra are much narrower than the Cu spectra and easier to obtain. In agreement with the Cu NMR results, the spectrum at 25.93 T with three sharp peaks split by the quadrupolar interaction indicates a uniform magnetization, whereas the rich structure of the spectrum at 27.82 T points to a magnetic superlattice. If the transition were second order and there were no disorder, we would expect that the spectral shape changes continuously with the field. This is, however, not the case: The spectrum at 26.82 T is a superposition of two types of spectra, each of which is identical to the spectrum at 25.93 T or 27.82 T. Thus, the two phases coexist at this field. The volume fraction of the uniform phase, given by the intensity of the sharp central peak, continuously changes in the field range 26.4 to 27 T without hysteresis (Fig. 4C). Such behavior can be explained if the transition is first order but broadened by some disorder in the sample, which enables nucleation of the magnetic superlattice with small domain size. Interestingly enough, this is consistent with the Landau-Lifshitz theory according to which the transition is expected to be first order when the ordering wave-vector does not lie on a high symmetry point; the ordering wave-vector $k = (\pi/a, \pi/2a)$ for the rhomboid cell is not a high symmetry point (*27*).

Quantum melting of various ordered structures − for example, vortex lattices in superconductors and charge or orbital order in perovskite transition metal oxides − has been attracting increasing attention. In many cases, the phase transition is controlled by chemical doping which also introduces disorder and increases complexity. Here we have a clean system with an exotic ordered structure, in which the phase transition can be tuned by a magnetic field. As the temperature is increased, we also expect a melting transition, which should be discontinuous for the same reasons discussed for the field-induced zero-temperature transition (*27*). This remains to be studied: at present we only know that the $^{11}$B NMR spectrum at 1.5 K shows no signature of superstructure.

**Supporting Online Material**
www.sciencemag.org
Supporting online text
Fig. S1

Table S1, Table S2

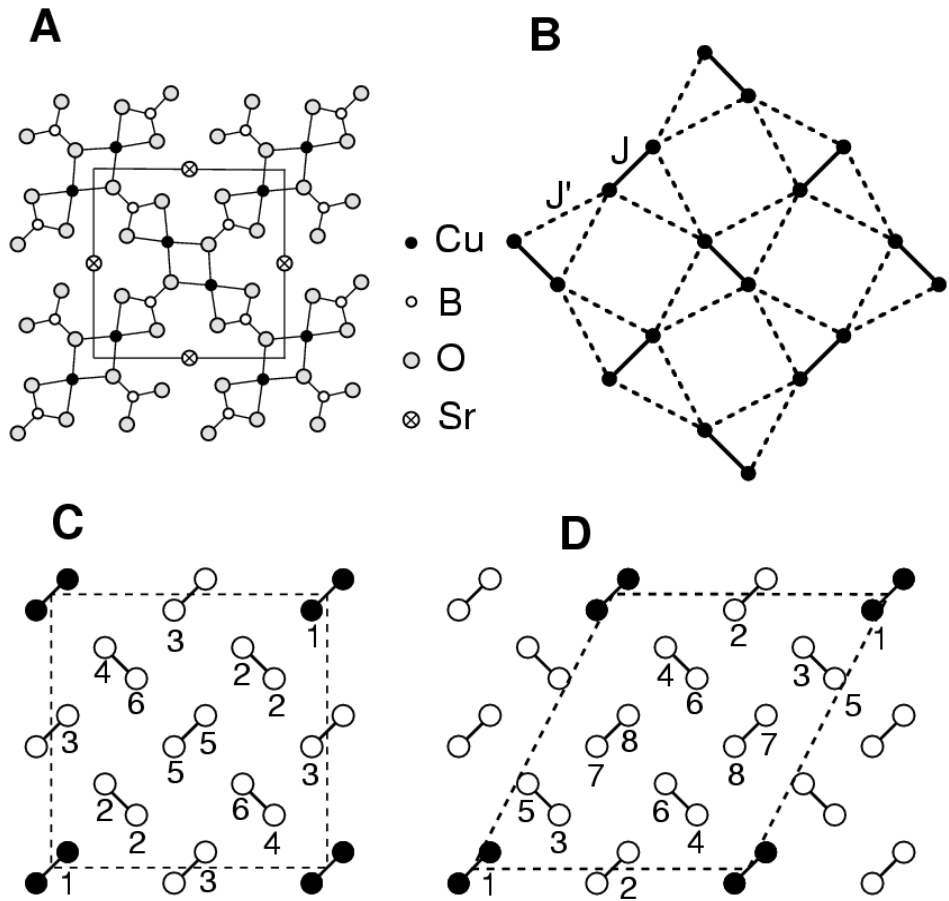

Fig.1. (**A**) The schematic structure of SrCu$_2$(BO$_3$)$_2$ viewed along the *c*-axis. The box shows a unit cell projected on the *ab*-plane. The structure consists of a stack of alternating Cu-B-O magnetic layers and nonmagnetic Sr layers. See (*9*) for details. (**B**) The Shastry-Sutherland spin model with the intra-dimer exchange *J* and the inter-dimer exchange *J'*, which is topologically equivalent to the Cu network within a layer of SrCu$_2$(BO$_3$)$_2$. (**C**) and (**D**) Proposed configurations of triplets for the 1/8 magnetization plateau phase (*21*). Solid circles denote the triplet dimers; numbers distinguish inequivalent Cu sites

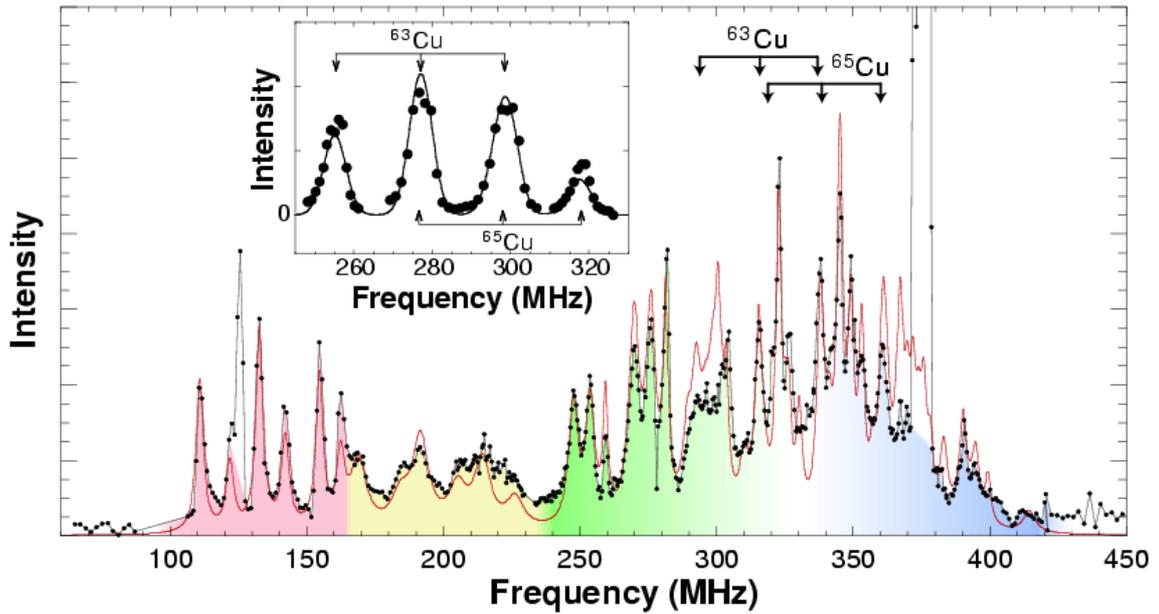

Fig 2. Cu NMR spectra (black dots) at 35 mK obtained by using a 20MW resistive magnet at the Grenoble High Magnetic Field Laboratory with a dilution refrigerator. The integrated intensity of Cu spin-echo signal is plotted at discrete frequencies. The inset shows the spectrum at $H_0$=26T. The line is a fit to Eq. 2 (see text). The arrows indicate the peak positions of the fitting. The main panel shows the spectrum at $H_0$=27.6 T. The red line is a fit by the sum of contribution from 11 distinct sites, each of which is represented by Eq. 2. The parameter values are listed in table S1. The arrows indicate the resonance frequencies for $H_n$=0. The strong signal in the range 371 to 379MHz comes from $^{11}$B nuclei; the peak at 125.6 MHz is due to $^{10}$B nuclei. The sensitivity of the NMR spectrometer was calibrated using $^{11}$B signal at 2 K at several frequencies. Because of unavoidable local variation of the sensitivity, the accuracy of the intensity (vertical) scale is estimated to be typically ±20%.

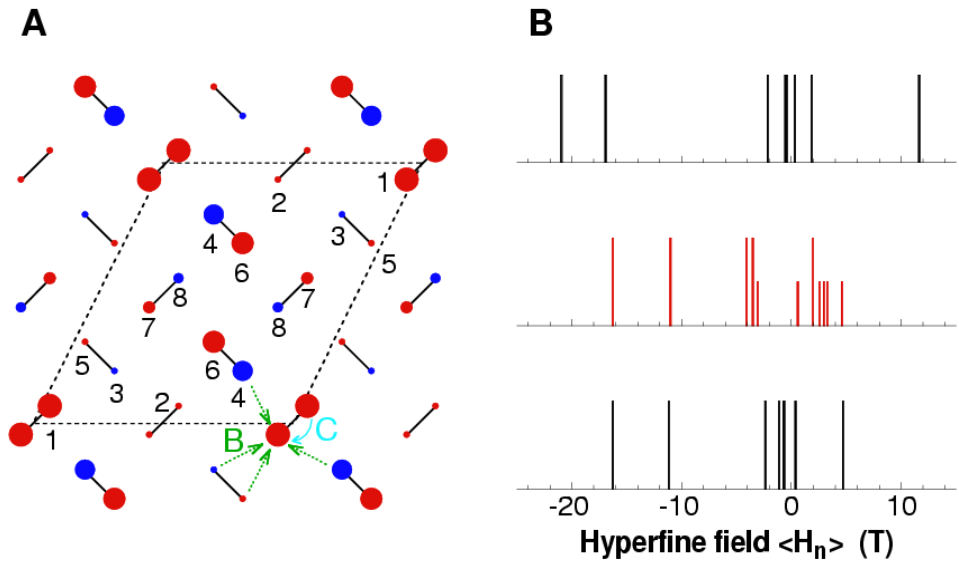

Fig. 3. (**A**) Magnetization profile obtained as described in the text. Red circles indicate positive $\langle S_z \rangle$, blue circles indicate negative $\langle S_z \rangle$, and circle size represents the magnitude of Red circles indicate positive $\langle S_z \rangle$. The calculated values for Red circles indicate positive $\langle S_z \rangle$ are 0.386 (site1), 0.010 (site 2), −0.0063 (site 3), −0.215 (site 4), 0.0073 (site 5), 0.312 (site 6), 0.040 (site 7), and −0.035 (site 8). (**B**) Histogram of the hyperfine field Red circles indicate positive $\langle H_n \rangle$. The middle panel (red) shows the results of the fitting of the NMR spectrum (Fig. 2, red line, and table S1). Long lines indicate that the population of the site is 1/8; short lines indicate that the population of the site is 1/16. The top panel is obtained from the theoretical magnetization profile in (**A**) assuming only the on-site hyperfine coupling $A=A_c=-23.8$ T/$\mu_B$ as described in the text. Generally there are also transferred hyperfine couplings to neighboring spins, denoted $B$ and $C$ in (**A**), but these are much smaller than the on-site coupling and are difficult to estimate. The bottom panel is obtained from the theoretical profile in (**A**) assuming $B=-1.04$, $C=-2.45$ T/$\mu_B$. (The previous low-field data (*12*) put the constraint $A=A_c-4B-C$.) The improved agreement with experiments, however, is only indicative.

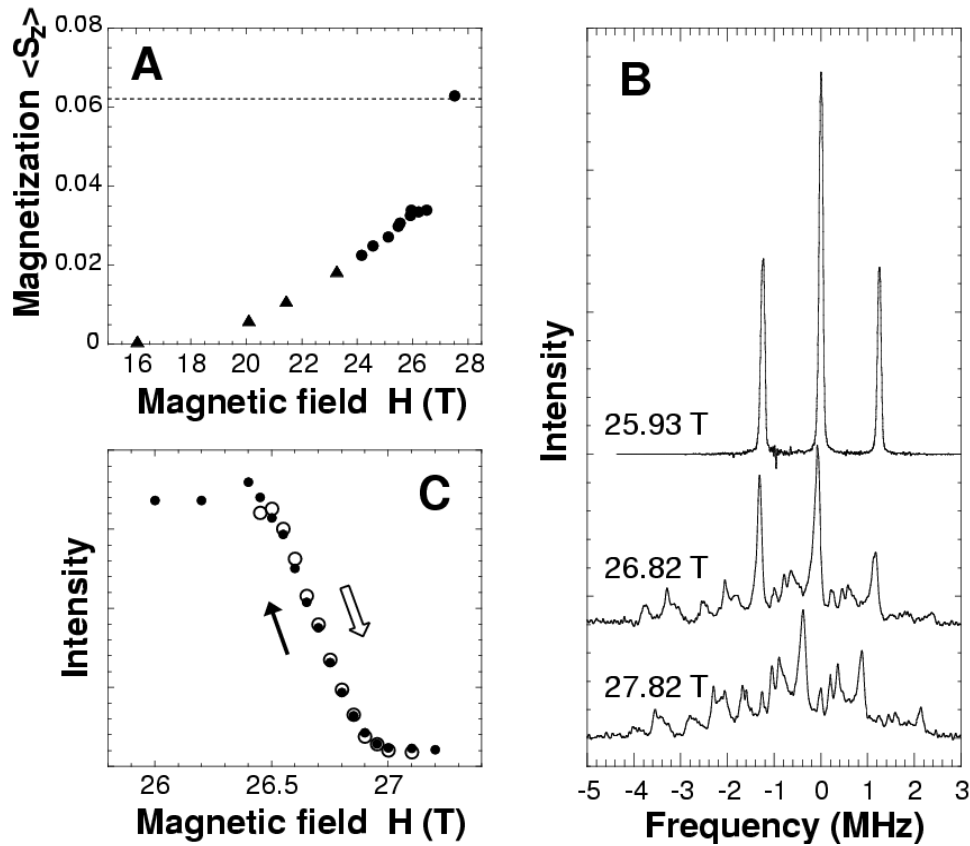

Fig. 4. (**A**) Field dependence of the average magnetization obtained from the Cu NMR spectra. Circles indicate data obtained at 35 mK; triangles denote data obtained at 0.5 K. The results below 26 T correspond to the uniform phase where the distribution of $H_n$ has a single peak. The horizontal line at $\langle S_z \rangle = 0.0625$ shows 1/8 of the saturation. (**B**) $^{11}$B NMR spectra at different magnetic fields obtained from summation of the Fourier-transformed spectra of the spin-echo signal at discrete frequencies. The frequency is measured from the unshifted position ($H_n=0$) for the quadrupole split central line. (**C**) Field dependence of the intensity of the $^{11}$B NMR central line of the uniform phase spectra measured with increasing (open circles) and decreasing (solid dots) field.

# Supporting online material

1) Magnetization on a finite system:

On a finite system with periodic boundary conditions, the ground state degeneracy is lower than 8-fold. This can be understood in the following way: The 8 states that would correspond to the 8-fold degenerate ground state of the infinite system are mixed by the boundary conditions, and there are 8 linear combinations. Some of them can remain degenerate, while the others would collapse onto the ground state only in the thermodynamic limit. For the square (rhomboid) 16-site cluster, the ground state is 4-fold (2-fold) degenerate, and no physically relevant magnetization can be defined before the degeneracy is completely lifted. Besides, the partial lifting of the degeneracy due to the boundary conditions is not physically satisfactory since it is clearly a finite-size effect. So a physically relevant mechanism has to be included to overcome this irrelevant source of degeneracy lifting and impose a physical effect.

2) Values of spin-phonon coupling and elastic constants:

No direct information is available on the parameters in Eq. 3, and we have taken values typical of other oxides:
- The spin-phonon coupling constants $a$ and $a'$ correspond to the exponent of the power law according to which the exchange integrals decrease with distance. This dependence was measured in $La_2CuO_4$ by applying pressure, giving the exponent between 6 and 7 ($1$). So we have taken a value of 7 for both $a$ and $a'$.
- The parameters $K$ and $K'$ are only effective parameters that cannot be directly matched to the phonon dispersion - an appropriate model of elastic constants should in any case include springs between all nearest neighbors. But this would not improve the calculation because the precise relationship between the actual position of all atoms and the superexchange constants between $Cu^{2+}$ spins is not known. Now the order of magnitude of $K$ and $K'$ is expected to be the same in all oxides, and for $CuGeO_3$, we have checked that relative displacements of less than a percent - as observed experimentally in the spin-Peierls phase ($2$) - require values of $K$ at least as large as 20000 Kelvin ($3$). This is the value we have taken for both $K$ and $K'$.

3) Order of the transition:

According to Landau theory, a transition can be second order if two necessary criteria are fulfilled, see e.g. ($4$):
- The so-called Landau criterion, which requires that no third-order invariant is present in the expansion of the free-energy in terms of order parameters.
- The so-called Lifshitz criterion, which requires in the case of a translational symmetry breaking that the ordering wave-vector be at a high symmetry point.
The combination of these criteria was for instance used to predict which surface reconstructions could be second order ($5$), and an ordering wave-vector equal to ($\boldsymbol{p}/a, \boldsymbol{p}/2a$) is excluded for a square lattice.

In the present case, assuming that:
- The 2D character is dominant,
- Landau theory can be applied at zero temperature, the role of the free energy as a function of

temperature being played by the ground state energy as a function of magnetic field, the transition is predicted to be first order because the Lifshitz criterion is violated.

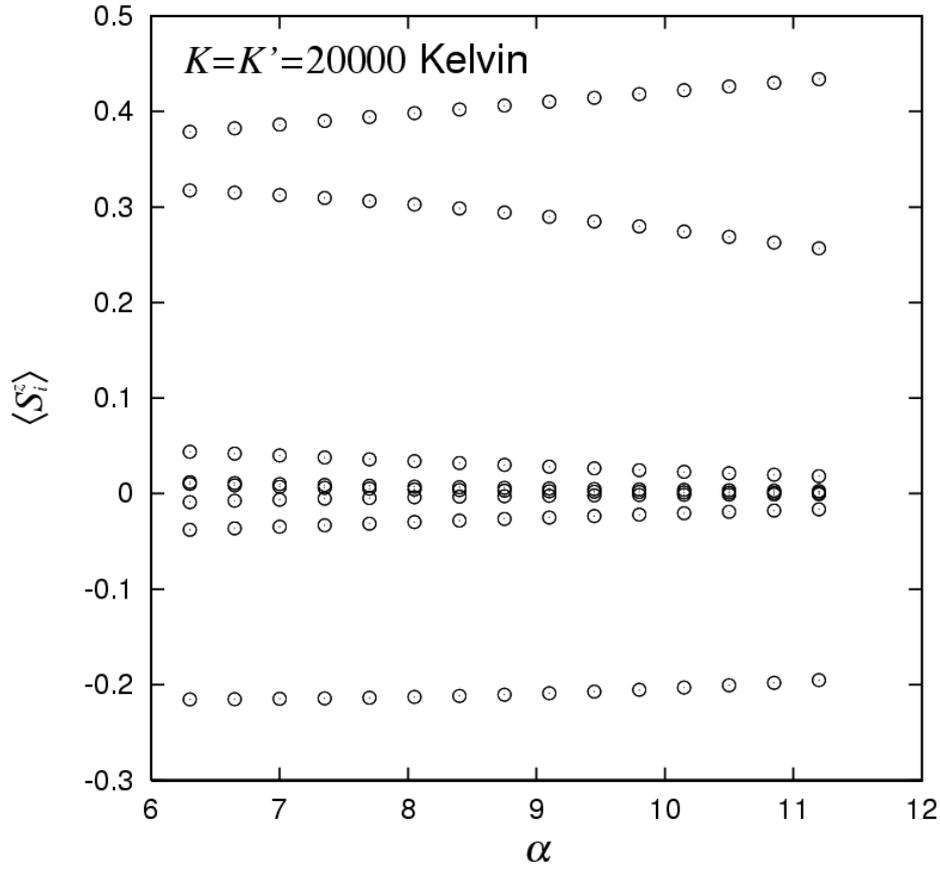

Fig. S1. The magnetization pattern as a function of the spin-phonon coupling *a* for a finite system ($J$ = 85 K, $J'$ = 54 K, $K = K'$ = 20000 K and *a* = *a'*). While the exponent observed in $La_2CuO_4$ is between 6 and 7 (*1*), the theory of superexchange combined with empirical dependences of hopping integrals (*6*) predicts that this exponent could be significantly larger. Therefore, we have calculated the dependence of the magnetization on *a* (=*a'* ) in the range from 6 to 12. The magnetization depends on *a* only weakly with smooth behavior. The main qualitative features – two sites with large positive magnetization, one site with large negative magnetization and others with small magnetization – remain over the whole range. The results provide convincing evidence that these features will be preserved in the limit of an infinite system, and for an infinitesimal spin-phonon coupling.

**Table S1. Parameters used to reproduce the Cu NMR spectrum at 27.6 T.**

| Site | $\langle H_n \rangle$ (T) | FWHM (T) | Intensity | $^{63}\nu_c$ (MHz) |
|------|---------------------------|----------|-----------|---------------------|
| 1    | −16.23                    | 0.28     | 1/8       | 21.9                |
| 2    | −11.01                    | 0.64     | 1/8       | 22.7                |
| 3    | −4.05                     | 0.35     | 1/8       | 22.5                |
| 4    | −3.51                     | 0.35     | 1/8       | 22.5                |
| 5    | −3.04                     | 0.18     | 1/16      | 22.1                |
| 6    | 0.60                      | 0.18     | 1/16      | 22.1                |
| 7    | 1.97                      | 0.27     | 1/8       | 22.7                |
| 8    | 2.61                      | 0.18     | 1/16      | 22.1                |
| 9    | 2.95                      | 0.21     | 1/16      | 22.7                |
| 10   | 3.30                      | 0.21     | 1/16      | 22.8                |
| 11   | 4.61                      | 0.44     | 1/16      | 22.1                |

**Table S2. Lengths of nearest-neighbor bonds corresponding to Fig. 3A.**

| Nearest-neighbor pair | 1-1 | 2-2 | 3-5 | 4-6 | 7-8 |
|---|---|---|---|---|---|
| Bond length (Å) | 2.9154 | 2.8890 | 2.8873 | 2.8958 | 2.8874 |